# A SIMPLE MODEL OF ENERGY EXPENDITURE IN HUMAN LOCOMOTION

*F. Romeo*

*Dipartimento di Fisica "E. R. Caianiello", Università degli Studi di Salerno*

*I-84081 Baronissi (SA) , Italy*

**ABSTRACT**

A simple harmonic oscillator model is proposed to describe the mechanical power involved in human locomotion. In this framework, by taking into account the anthropometric parameters of a standard individual, we are able to calculate the speed-power curves in human walking. The proposed model accounts for the well known Margaria's law in which the cost of the human running (independent from the speed) is fixed to *1 Kcal/(Kg Km)*. The model includes the effects of a gentle slope (either positive or negative) and the effect due to the mechanical response of the walking surface. The model results obtained in the presence of a slope are in qualitative agreement with the experimental data obtained by A. Leonardi *et al.*

## 1. INTRODUCTION

Several studies have been done to clarify the biomechanical, metabolic and energetic aspects involved in human and bipedal locomotion [1-6]. Due to the tremendous implications in different fields of our life (health, sport, medicine, biology, robotics [7, 8], etc), the research on human motion can be considered as a multidisciplinary domain, where different competences are needed. Due to this, the study of the energetic aspects related to human gait could be considered as a formidable task. Nevertheless, elementary physics can be adopted to obtain several useful indications about energy expenditure in human walking. Following the spring-mass model proposed by R. M. Alexander [9-12], the extremely low energetic cost of human walking and running is strongly related to the role of the tendons acting as elastic energy storage elements. For example, the elastic energy stored in the right (left) knee bending act, due to the falling of the right (left) foot to the ground, is released at the beginning of a new step. For instance, one can notice that the mentioned mechanism becomes more efficient as the speed of gait increases. In this way, in the range of running velocity *4-6 m/s*, the energy expenditure becomes insensitive to the speed and quite surprisingly saturates to a value close to *1 Kcal/(Kg Km)*, as prescribed by the Margaria's law [13, 14].

In this work we shall derive the Margaria's law by means of a mechanical model consisting of a forced harmonic oscillator. In this context, we can understand the experimental behaviour of a typical speed-power curve by invoking the presence of a point of inflection, which mimics a saturation of the energy expenditure to *1 Kcal/(Kg Km)* for velocities below a threshold value, which is not accessible to human runners. The effect of elastic energy storage of the ground-foot-tendon system is also included in the model by means of a coefficient of restitution $\alpha$, measuring the ratio between the speed of the impacting foot and its exiting speed from the impact, the latter velocity being evaluated at the beginning of the leg swinging phase. In conclusion, the effects of a moderate positive or negative slope are considered.

## 2. THE MODEL

We describe the gait of an individual characterized by a total mass $m$ whose leg has mass, length and momentum of inertia respectively equal to $\mu$, $\lambda$ and $I$. The leg is schematized as a single linear homogeneous element. The slope $\gamma$ of the ground (both for positive and negative values) is taken into account. The walking model we shall consider consists of three phase (see Fig.1): *1)* rising of the centre of mass; *2)* leg swinging of amplitude $\theta$; *3)* lowering of the centre of mass. In the walking action, when displacing the centre of mass in the walking direction of a quantity $2\lambda\sin(\theta/2)$, there is at least one foot in contact with the walking surface. In order to fully characterize the dynamics of the system, we need to write the dynamical equation of the swinging leg subjected to an external torque. Therefore, by defining $\beta$ as the angle formed by the swinging leg with respect to the direction of the gravitational force, one can write (for small $\beta$):

$$\frac{d^2\beta}{dt^2}+\frac{\mu g\lambda}{2I}\beta=\frac{M_{ex}}{I}, \tag{1}$$

where $M_{ex}$ is the applied torque required to the muscles to obtain the leg motion. Eq. (1) can be solved by imposing the initial conditions $\beta(0)=-\theta/2+\gamma$ and $\beta'(0)=\omega_0$. Proceeding as we said, by defining $\omega=\sqrt{\frac{\mu g\lambda}{2I}}$, we obtain the following expression:

$$\beta(t)=-\left(\frac{\theta}{2}+\frac{M_{ex}}{I\omega^2}\right)\cos(\omega t)+\gamma\cos(\omega t)+\frac{\omega_0}{\omega}\sin(\omega t)+\frac{M_{ex}}{I\omega^2}. \tag{2}$$

We are now interested in the computation of the value of the external torque $M_{ex}$ we have to apply to obtain a given mean velocity $u$. Observing that in the time interval $\tau$, which elapses in swinging the leg between the angular positions $[-\theta/2+\gamma,\theta/2+\gamma]$, the system displace its centre of mass of a quantity $2\lambda\sin(\theta/2)$, we get $\tau\cong\frac{2\lambda\sin(\theta/2)}{u}$. Solving now Eq.(2) for $M_{ex}$, imposing $\beta(\tau)=\theta/2+\gamma$, we obtain the following relation:

$$M_{ex}=\frac{I\omega^2\theta}{2}\left(\frac{1+\cos(\omega\tau)}{1-\cos(\omega\tau)}\right)+\gamma I\omega^2-\frac{\omega_0}{\omega}\frac{\sin(\omega\tau)}{1-\cos(\omega\tau)}, \tag{3}$$

which is the constant torque we have to apply to the leg in order to complete a step at a mean velocity $u$. The average mechanical power $\langle w_s\rangle$ involved in swinging the leg can be easily computed by means of the following relation coming from the standard definition:

$$\langle w_s\rangle=\frac{M_{ex}}{\tau}\int_{-\theta/2+\gamma}^{\theta/2+\gamma}d\vartheta=\frac{M_{ex}\theta u}{2\lambda\sin(\theta/2)}. \tag{4}$$

Eq. (4) can be now recast taking into account both Eq. (3) and the explicit value of the momentum of inertia $I=\frac{\mu\lambda^2}{12}+\frac{\mu\lambda^2}{4}=\frac{\mu\lambda^2}{3}$ as follows:

$$\langle w_s \rangle = \frac{\mu g u \theta^2}{8\sin(\theta/2)}\left(\frac{1+\cos(\omega\tau)}{1-\cos(\omega\tau)}\right) - \frac{\omega_0 \theta \mu u \sqrt{g\lambda}}{2\sqrt{6}\sin(\theta/2)}\frac{\sin(\omega\tau)}{1-\cos(\omega\tau)} + \frac{\gamma g u \theta \mu}{4\sin(\theta/2)}. \tag{5}$$

Apart from the above contribution to the mechanical power involved in human walking, we also need to consider the power involved in the process of rising the centre of mass of the system. This contribution, which in the case of null slope can be calculated as

$$\langle w_{cm} \rangle = \frac{mg\,\Delta h}{\tau/2} = \frac{mgu(1-\cos(\theta/2))}{\sin(\theta/2)}, \tag{6a}$$

becomes

$$\langle w_{cm} \rangle = \frac{mgu(1-\cos(\theta/2+\gamma))\theta}{\sin(\theta/2)(\theta+2\gamma)}, \tag{6b}$$

for arbitrary but gentle slope. Finally the total mechanical power is obtained as the sum $\langle w \rangle = \langle w_s \rangle + \langle w_{cm} \rangle$. Nevertheless, the average total power presents an unwanted dependence on the angle $\theta$. This problem can be overcome following J. E. A. Bertram and A. Ruina [15] which find an explanation to the empirical relation linking the amplitude of the step to the mean velocity as the result of the constrained minimization of the $O_2$ consumption. This empirical relation in our notation reads as follows:

$$\lambda \sin(\theta/2) \propto u^{1-\varepsilon}, \tag{7}$$

where $\varepsilon = 0.58$. It is worth to notice that the previous relation is not trivially related to the minimization of the function $\langle w \rangle$ with respect to $\theta$, but is strongly determined by biological conditions. Substituting Eq. (7) in the expression for the mechanical power $\langle w \rangle$, we can write the energy expenditure $W$ involved in human walking in units *Kcal/(Kg Km)* as follows:

$$W = \frac{\langle w \rangle}{(4.186)\, m \eta u}\frac{Kcal}{Kg\,Km}, \tag{8}$$

where the numerical values which appears in Eq. (8) are in S.I. units, while the coefficient $\eta$ represents the muscles efficiency, which is a quantity strongly affected both by climatic parameters (temperature, humidity, etc.) and by metabolic ones.

## 3. RESULTS

In order to compare our results with the experimental data, we analyse Eq. (8) by fixing $g = 9.8\, m/s^2$ and the remaining anthropometric parameters as follows: $m = 70Kg$ ; $\mu = 17.5Kg$ ; $\lambda = 0.9$ . In addition, we fix a coefficient of proportionality, let say *A*, in Eq. (7) to be $0.17\,(m/s)^{\varepsilon}\, s$ . It is worth to notice that the constant *A*, which in the real world depends on the individual features, is not arbitrary, but can be theoretically assigned by imposing a condition on the maximum step length obtained at the maximum mean velocity. As a final assumption, we set $\omega_0 = \alpha \frac{u}{\lambda}$ , where the non-dimensional factor $\alpha$ plays a role which will be clear in the following discussion.

In Fig.2, assuming a null slope, we presents the energy expenditure *W* as a function of the mean velocity *u* (m/s) for four different values of the parameter $\alpha$ (from bottom to top $\alpha = 1.12$, $\alpha = 1.1$, $\alpha = 1$, $\alpha = 0.9$), setting $\eta = 0.30$. As can be noticed, a saturation behaviour is obtained for $\alpha = 1.1$, according to the well known Margaria's law.

In Fig. 3a we presents the effect of different muscles efficiency ranging from 0.25 to 0.35 leaving the remaining parameters as in Fig. 2. The previous features related to the plateau of the curves are preserved even though the saturation value varies, though remaining close to *1 Kcal/(Kg Km).*

The behaviour presented above can be associated to the formation of an inflection point in the *W* vs *u* curves, as shown in Fig.3b. In fact, in Fig. 3b an extended range of velocities is used in order to emphasize the analytic behaviour responsible for the saturation observed in the experimental data. The condition for the formation of the plateau is related to the parameter $\alpha$, as shown in Fig. 2. This parameter can be altered in several ways including the nature of the ground [1, 2] (rigid or absorbing ground) or the muscles status. Indeed, when walking under tired, a loss of efficiency in the elastic energy storage mechanism of the system muscle-tendon appears. Finally, motivated by recent experimental works by A. Leonardi *et al.*, in Fig. 4 we plot the energy expenditure as a function of a gentle slope $\arctan(\gamma) \approx \gamma$ for different values of $\alpha$, mimicking different ground responses. As shown in Fig. 4, a linear behaviour is obtained in the slope in agreement with ref.

[16]. The additional energy cost associated with $\gamma$ can be extracted from Eq. (8) by Taylor's expansion to the first order in $\gamma$. Following this procedure, we get the following formula:

$$W_\gamma = \frac{\gamma g}{\eta(4.186)}\left(\frac{\mu}{m}\frac{\theta}{4\sin(\theta/2)}+\frac{1}{2}\right), \qquad (9)$$

which can be simplified by means of the approximations $\sin(\theta) \approx \theta$ and $m \approx 4\mu$. In this way, provided that $\eta = 0.30$, for a slope of $\pm 0.05$ the additional energy cost assumes the numerical value $\pm 0.244$ *Kcal/(Kg Km)*, while for a slope $\pm 0.025$ the additional cost becomes one half of the previous value.

## 4. CONCLUSIONS

We presented a model accounting for the energy expenditure in human motion. The well known Margaria's law is recovered and the saturation behaviour observed in the experimental *W* vs *u* curves is explained by imposing the presence of an inflection point, controlled by the parameter $\alpha$, in those same curves. The supplementary cost due to a slope $\gamma$ is analyzed and the additional effect related to the mechanical properties of the ground is also included. Concerning the effect of the slope, the supplementary cost is roughly found to be equal to $\gamma \times 4.88$ *Kcal/(Kg Km)*, where the muscles efficiency is set to 0.30, $\gamma$ being expressed in radians. Even though the presented model is rigorously valid only for walking and for race walking, we expect the obtained results to be valid also in the case of running. In addition, the results shown in Fig. 4 are in qualitative agreement with experimental results presented in Ref. [16]. In order to obtain also a quantitative agreement, we should be able to fit the experimental data taking into account the mass and $\eta$ dependence present in Eq. (9), leaving the parameter $\alpha$ (characterizing the ground mechanical properties) as a free parameter to be determined by means of linear regression. Good quantitative agreement is expected only if the ground properties do not affect too much the walking mechanics.

**ACKNOLEDGMENTS**

The author would like to thank Roberto De Luca for several enlightening discussions on the topic of this work.

**FIGURES**

**FIG.1**

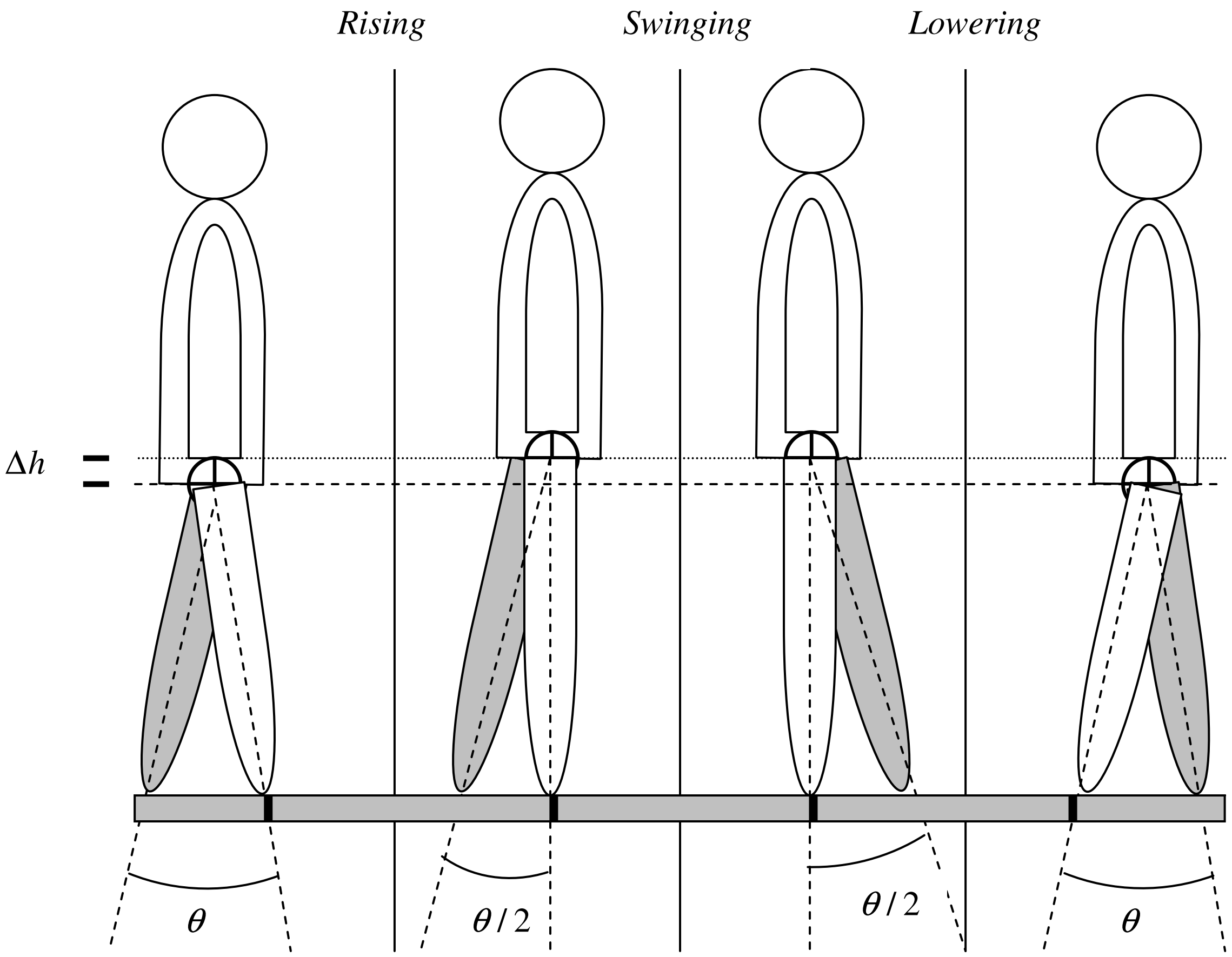

**FIG.2**

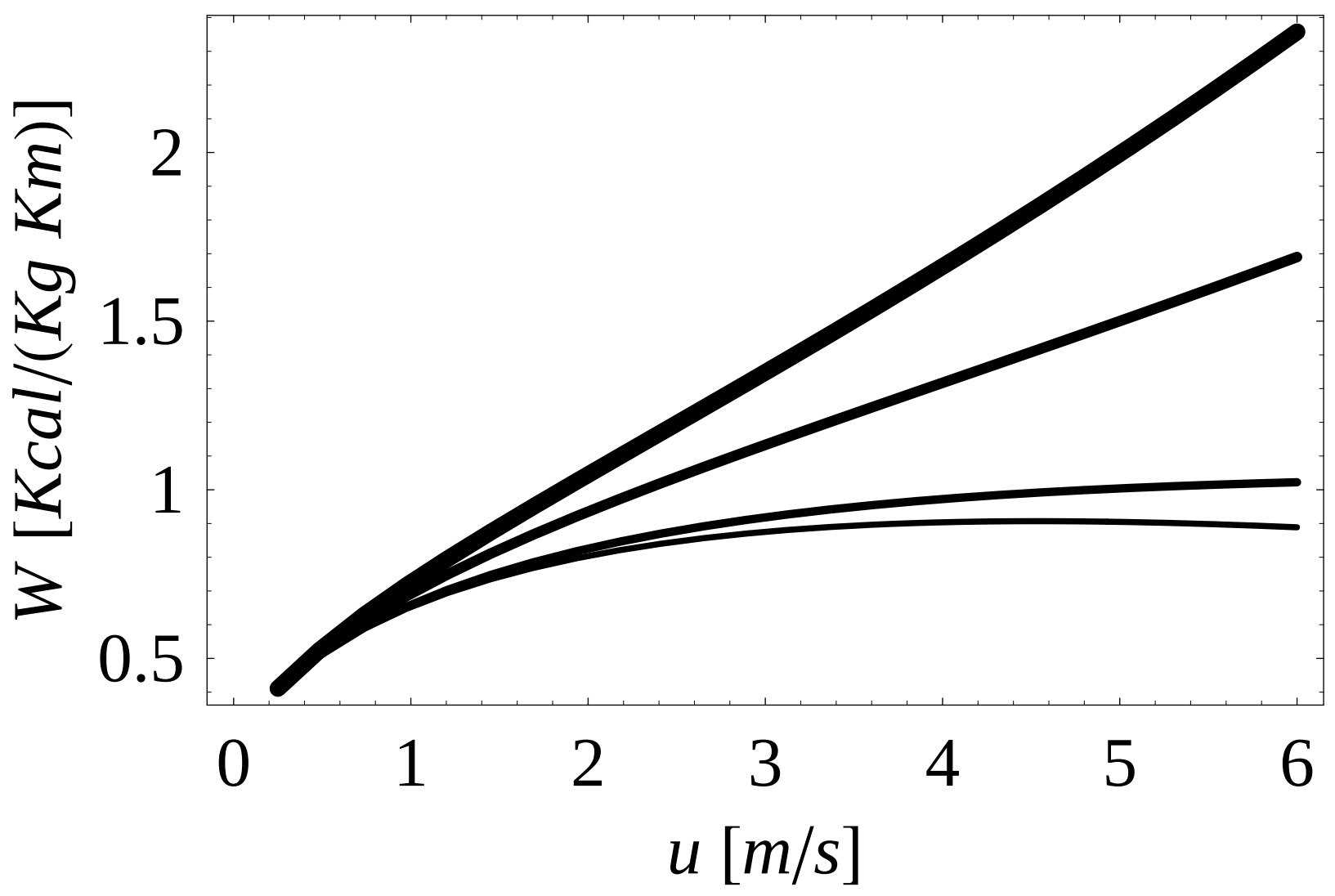

**FIG.3a**

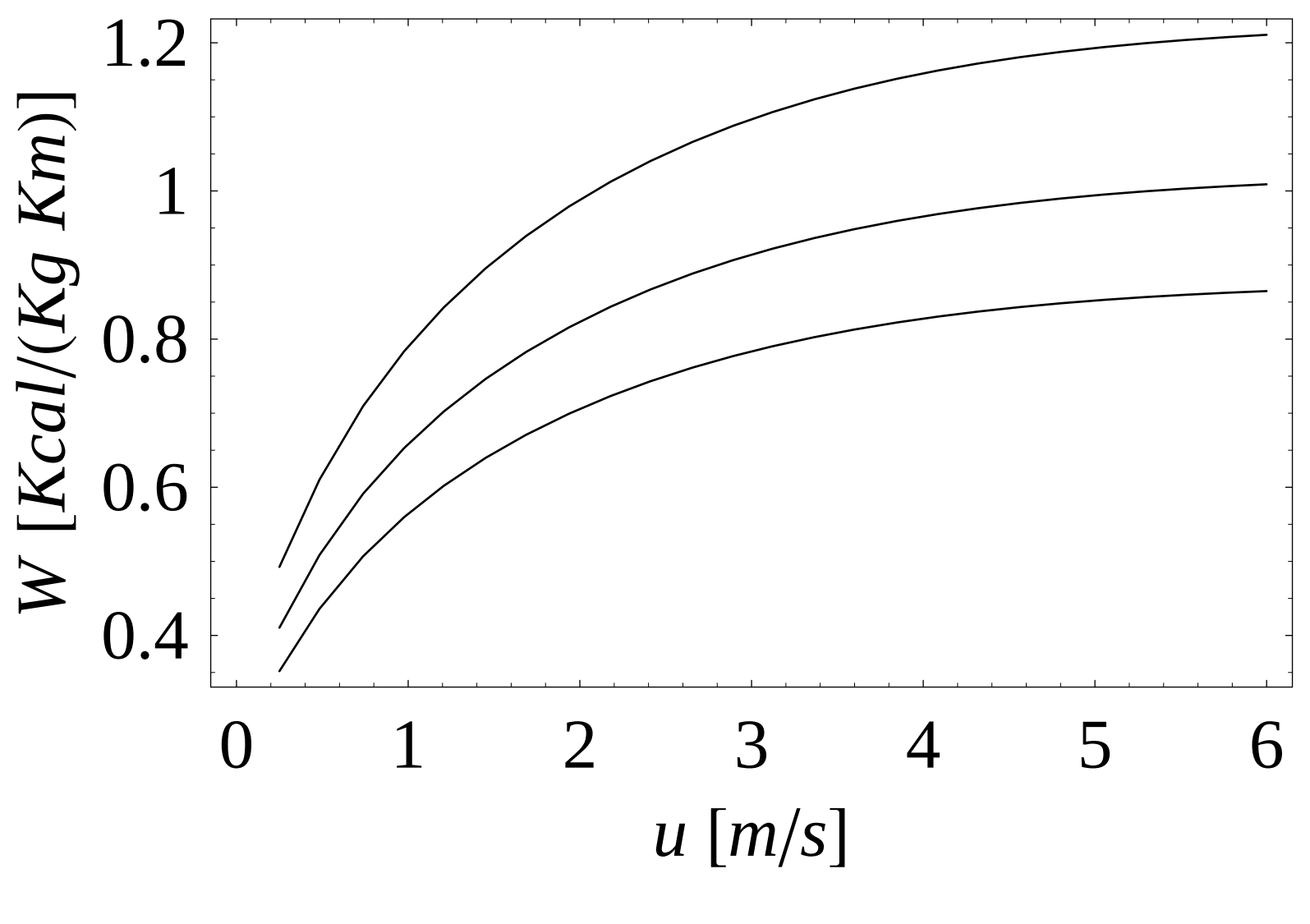

**FIG.3b**

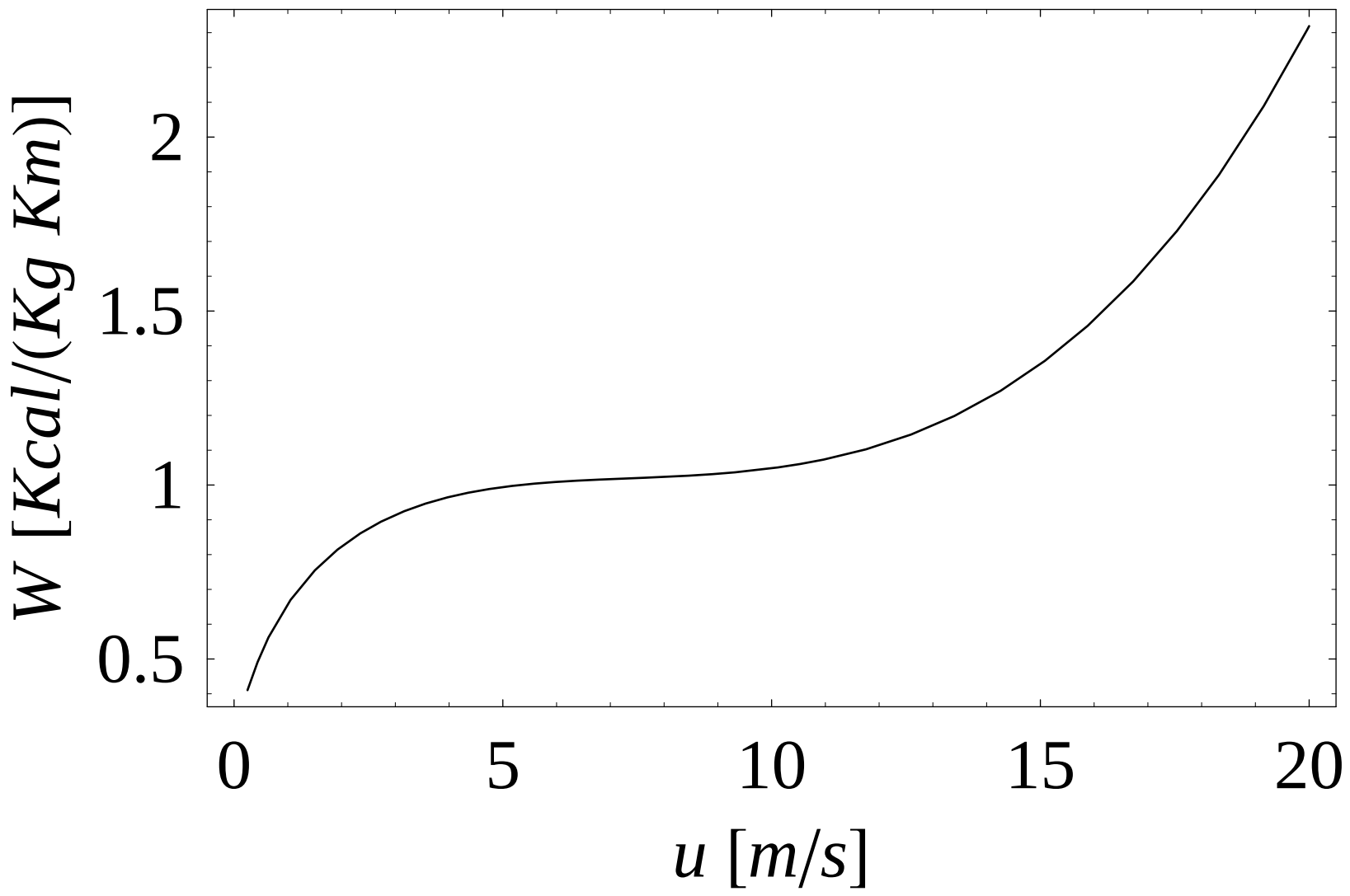

**FIG.4**

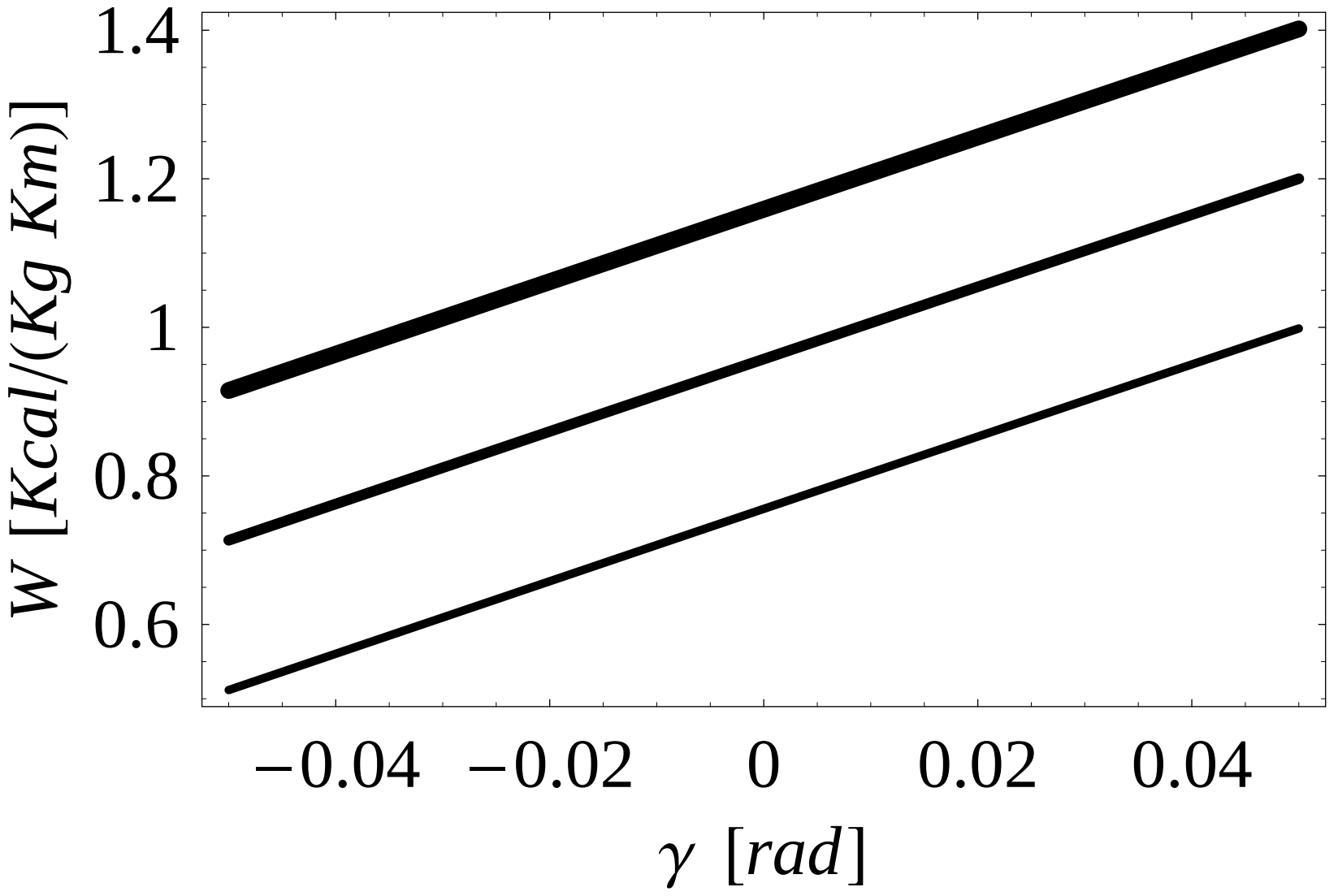

## FIGURE CAPTIONS

### FIG.1

A schematic representation of the three phases involved in human walking as described in text for a null slope.

### FIG.2

Energy expenditure of the human locomotion as a function of the mean velocity *u (m/s)*, obtained by choosing the parameters of the model as follows: $\gamma = 0$; $m = 70Kg$ ; $\mu = 17.5Kg$ ; $\lambda = 0.9$ ; $\eta = 0.30$ ; $A = 0.17\left(m/s\right)^{\varepsilon} s$ . The curves refers to different values of the parameter $\alpha$ , respectively from bottom to top: $\alpha = 1.12$ , $\alpha = 1.1$ , $\alpha = 1$ , $\alpha = 0.9$ .

### FIG.3a

Energy expenditure of the human locomotion as a function of the mean velocity *u (m/s)*, obtained by choosing the parameters of the model as follows: $\gamma = 0$; $m = 70Kg$ ; $\mu = 17.5Kg$ ; $\lambda = 0.9$ ; $\alpha = 1.1$; $A = 0.17\left(m/s\right)^{\varepsilon} s$ . The curves refers to different values of the parameter $\eta$, respectively from bottom to top: $\eta = 0.35$ , $\eta = 0.30$ , $\eta = 0.25$ .

### FIG.3b

Energy expenditure of the human locomotion as a function of the mean velocity *u (m/s)*, obtained by choosing the parameters of the model as follows: $\gamma = 0$; $m = 70Kg$ ; $\mu = 17.5Kg$ ; $\lambda = 0.9$ ; $\alpha = 1.1$; $A = 0.17\left(m/s\right)^{\varepsilon} s$ ; $\eta = 0.30$ . This curve, corresponding to the middle curve in Fig. 3a, shows clearly the presence of an inflection point when the *u* range is increased.

### FIG.4

Energy expenditure of the human locomotion as a function of the slope $\gamma$, obtained by choosing the parameters of the model as follows: $u = 1.5 m/s$ ; $m = 70Kg$ ; $\mu = 17.5Kg$ ; $\lambda = 0.9$ ; $\eta = 0.30$ ;

$A = 0.17\,(m/s)^{\varepsilon}\,s$ . The curves refers to different values of the parameter $\alpha$ , respectively from bottom to top: $\alpha = 1.1$ , $\alpha = 0.8$ , $\alpha = 0.5$ .